\documentclass[twocolumn,showpacs,preprintnumbers,amsmah,amssymb]{revtex4}
\usepackage{graphicx}

\begin{document}

\title[]{Magnetic induction and domain walls in magnetic thin films at remanence}

\author{Florin Radu
\footnote{To whom correspondence should be addressed
(florin.radu@rub.de)},  Vincent Leiner, Kurt Westerholt, Hartmut
Zabel}

\address{Institut f\"{u}r
Experimentalphysik/Festk\"{o}rperphysik, Ruhr-Universit\"{a}t
Bochum, D-44780 Bochum, Germany}

\author{Jeffery McCord}

\address{Leibniz Institute for Solid State and Materials Research
Dresden, Helmholtzstrasse 20, D-01171 Dresden, Germany}

\author{Alexei Vorobiev,  Janos Major}
\address{Max-Planck-Institut   f\"ur Metallforschung,
Heisenbergstr. 3, D-70569 Stuttgart, Germany}

\author{David Jullien, Hubert Humblot,  Francis Tasset}
\address{Institut Laue-Langevin, F-38042 Grenoble Cedex 9,
France}

\begin{abstract}

Magnetic domain walls in thin films can be well analyzed using
polarized neutron reflectometry. Well defined streaks in the
off-specular spin-flip scattering maps are explained by neutron
refraction at perpendicular N\'{e}el walls. The position of the
streaks depends only on the magnetic induction within the domains,
whereas the intensity of the off-specular magnetic scattering
depends on the spin-flip probability at the domain walls and on
the average size of the magnetic domains. This effect is
fundamentally different and has to be clearly distinguished from
diffuse scattering originating from the size distribution of
magnetic domains. Polarized neutron reflectivity experiments were
carried out using a $^3$He gas spin-filter with a analyzing power
as high as 96\% and a neutron transmission of approx 35\%.
Furthermore, the off-specular magnetic scattering was enhanced by
using neutron resonance and neutron standing wave techniques.

\end{abstract}



\maketitle

\section*{Introduction}
Magnetic domain imaging at remanence and during magnetization
reversal has become an extremely important discipline in the area
of  magnetic thin films. For magnetoelectronic and spintronic
device applications, the domain state and switching behavior of
various vertically and laterally structured magnetic systems
requires detailed investigations in the space and time domain. The
experimental techniques available for domain imaging provide
either real space information, such as magnetic force microscopy
(MFM)\cite{rugar:1990,hartmann:1999}, Kerr microscopy
(KM)\cite{mccord:2003}, Lorentz microscopy
(LM)\cite{zweck:1997,kirk:1999,bruckner:2004}, polarized electron
emission microscopy (PEEM)\cite{fisher:1998,kuch:2003}, and
secondary electron microscopy with polarization analysis
(SEMPA)\cite{scheinfein:1990}, or reciprocal space information,
such as resonant soft x-ray magnetic small-angle scattering
(SAS)\cite{kortright:2001}. In any case, magnetic domain
information is obtained via magnetic stray fields emanating from
the sample (MFM) or via the local magnetic polarization. Most of
the methods presently available do not concentrate on the magnetic
domain walls, their thickness, and
their orientation.\\
In this paper we report on the observation of magnetic domain
walls, using polarized neutron reflectivity (PNR). PNR has proven
in the past to be an essential tool for the analysis of magnetic
thin films and heterostructures, including their interfaces and
domain structures\cite{fitzsimmons,ankner,prb2003}. Here we will
go one step further and show that magnetic domain walls can be
characterized via pronounced streaks in the off-specular spin-flip
(SF) scattering regime. Usually, off-specular SF neutron
reflectivity is symmetric (R$-+$ = R$+-$)  and is discussed in
terms of the Fourier transform of the magnetic domain distribution
in the film plane and/or in terms of interfacial magnetic
roughness~\cite{suzane2001, langridge2000, tope2001, lee2002,ktb}.
Off-specular asymmetric SF reflectivity (R$-+$$\ne$ R$+-$) can  be
observed, if the magnetic induction in the sample and the external
magnetic field to the sample are not
collinear\cite{ignatovich1978,felcher1995,korneev1996,
fredrikze1998,kruijs2000}. This leads to SF processes at the
sample/vacuum interface, which are always accompanied by a Zeeman
energy change of the neutrons. Consequently, the SF neutrons
follow a different path in the external applied field
(off-specular) than the non-spin flip (NSF) neutrons (specular).
The effect we report here has to be clearly distinguished from
either one of these cases. We observe a striking new and
asymmetric off-specular neutron reflectivity in the remanent state
of a Co film, which is bunched into pronounced streaks. We argue
that the asymmetry is caused by breaking the translational
symmetry for the neutrons travelling in the film plane, when
crossing domain walls with different magnetic potentials on either
side of the wall. Neutron scattering at Bloch walls in bulk
ferromagnets has been reported in the literature~\cite{schaerpf78,
schaerpf88, schaerpf89, podurets, peters}. In the present work, we
consider neutron scattering from domain walls in thin films with
an incident angle to the domain walls close to normal. Analysis of
the characteristic features of the off-specular SF scattering
reveals the magnetic induction of the film at remanence, the
average domain wall width,  the average
domain size, and domain wall angle.\\
Since neutron scattering at domain walls is weak, enhancement
factors have been employed, using neutron standing wave properties
below the critical angle of total reflection, and neutron
resonance conditions above the critical angle. A detailed
theoretical background and description of the experimental
conditions necessary to realize either Bright-Wigner resonances or
neutron standing waves in thin magnetic films, is provided in
Ref.~\cite{raduexp}. Furthermore, the method requires a full spin
analysis of the off-specular spin-flip intensity over a large
solid angle. This has been achieved with a supermirror in the
incident beam and a  $^3$He spin filter in the exit beam. The
polarized   gas was provided by TYREX 3rd generation  $^3$He
filling station recently commissioned at ILL~\cite{illrep}.

\section*{Experimental Results}
For the present experiment we used a 250~nm~thick Co film, grown
by rf-sputtering on a SiO$_{2}$ substrate. Due to exposure to air,
a 2.5~nm ~thick oxide layer is formed on top of the Co layer.
Growth conditions and substrate choice provided a polycrystalline
Co film.
\begin{figure}[!ht]
    \begin{center}
\includegraphics[clip=true,keepaspectratio=true,width=1\linewidth]{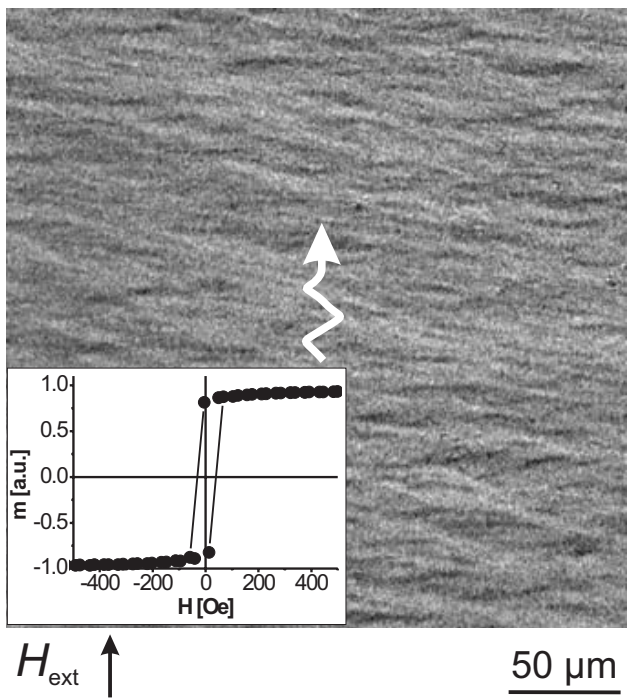}
    \end{center}
        \caption{\label{ripple}Ripple-like domain structure after applying a saturation field
of $H_{ext} = 2000$~Oe and returning to close to the remanent
state at $H_{ext} = 10$~Oe. The mean magnetization direction
together with the ripple like domain structure is indicated by the
curly arrow. The inset shows the hysteresis loop measure by MOKE}
\end{figure}
The magnetization reversal was characterized by the
magneto-optical Kerr effect (MOKE) and by wide field Kerr
microscopy in the longitudinal mode. MOKE hysteresis measurements
did not reveal any macroscopic anisotropy within the film plane. A
characteristic Kerr domain image is shown in Fig.~\ref{ripple},
taken after saturating the sample at +2000 Oe and reducing the
field to 10 Oe~\footnote{this field close to remanence was chosen
to be comparable to the guiding field during PNR measurements.}. A
strongly modulated low angle domain structure can be recognized,
which is due to the polycrystalline nature of the film combined
with the intrinsic magnetocrystalline anisotropy of Co. The
magnetization within the ripple  domains is partly tilted to the
left and to the right, and is essentially perpendicular to the
applied magnetic field (spin flop orientation).  The ripple can be
recognized  also in the hysteresis loop shown in the inset of
Fig.~\ref{ripple}, where a reduced remanence is observed. The
magnetization direction within the domains increasingly deviates
from the mean magnetization direction  when approaching the
coercive field, where the reversal of magnetization proceeds by
domain wall motion. At remanence, the deviation of the local
magnetization vector from the mean direction  can also be
recognized from the hysteresis loop shown in the inset of
Fig.~\ref{ripple}.

\begin{figure}[!ht]
\begin{center}
    \includegraphics[clip=true,keepaspectratio=true,width=1\linewidth]{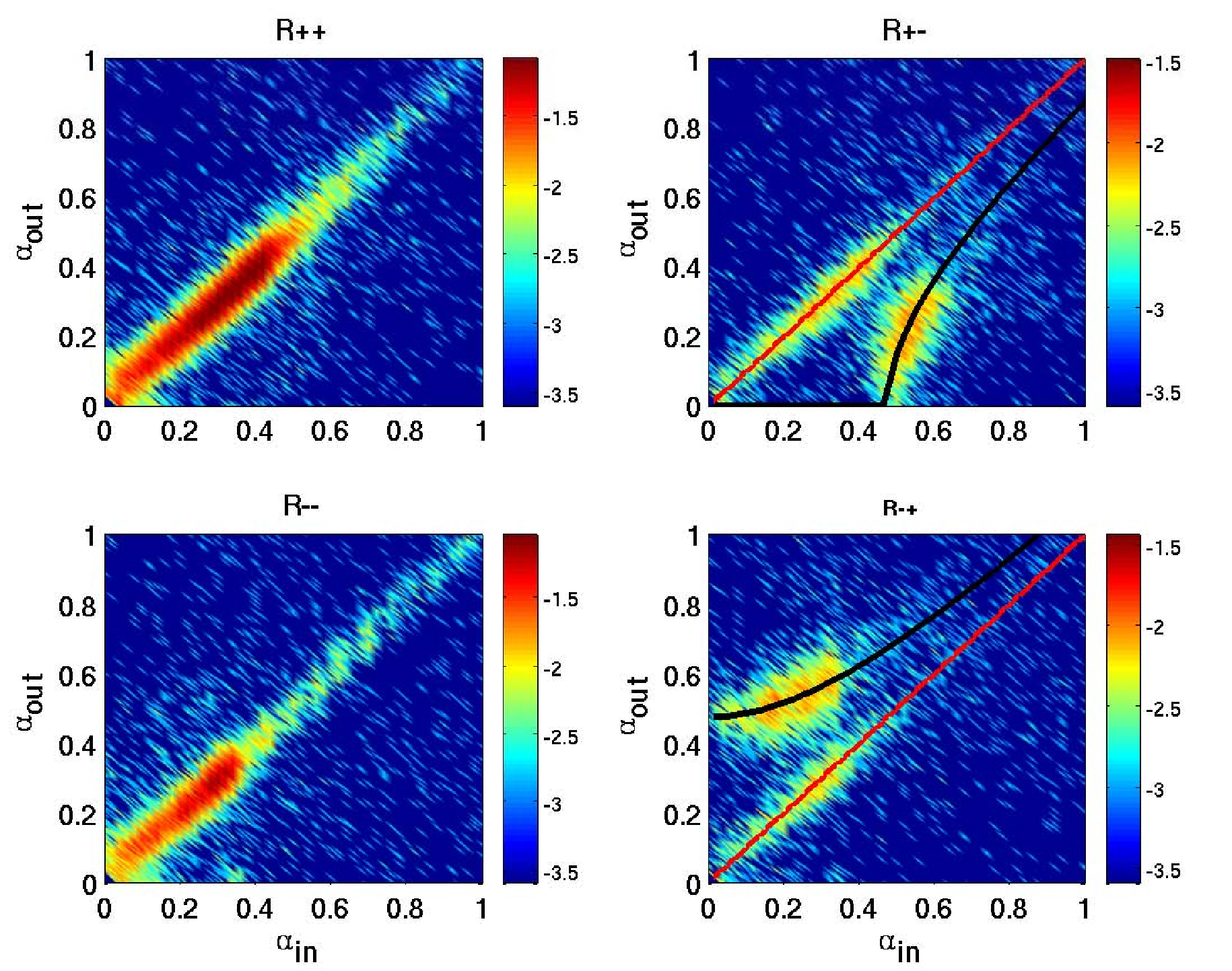}
    \end{center}
        \caption{\label{offspecEVA1}The left column shows intensity maps for the
        non spin-flip cross sections R$++$ (top) and R$--$ (bottom),
        the right column for the spin-flip
        cross sections R$+-$ (top) and R$-+$ (bottom). The maps were recorded close to remanence($\approx$~10~Oe).
        The intensities are plotted in terms of
        exit angles (y-axis) versus incident angle (x-axis).
        The streaks are due to reflection and refraction
        from perpendicular magnetic domain walls, as described in more detail in the text. The black lines
        are calculated off-specular spin-flip streaks. }
\end{figure}

We have carried out the neutron reflectivity measurements using
the ADAM and  EVA reflectometers at the Institut Laue-Langevin,
Grenoble, EU. The EVA reflectometer is used in combination with a
wide angle $^3$He spin filter analyzer. The half-life time of the
spin filter was 180~h and the analyzing power of the filter  was
96\% ,
 at the
beginning of the experiment, which is comparable to the best
performance of a supermirror analyzer. The neutron measurements
were carried out at 300 K and for two magnetic states of the
sample: saturation (not shown here) and remanence.
First the sample was saturated, then the field was reduced to zero
field (termed \textit{pristine state}).
The pristine  state is shown in Fig.~\ref{offspecEVA1}.

The diagonal intensity ridges stretching from the lower left to
the upper right corner are the specular NSF R$++$ and R$--$
reflectivities. The specular ridge is rich in features, confirming
standing waves and Kiessig fringes in the R$++$ map above the
critical edge for total reflection at the Co film, and standing
wave features in the R$--$ map below the critical edge of the
substrate. The situation is dramatically different for the
spin-flip(SF) maps R$+-$ and R$-+$. Aside from the specular
ridges, there is additional off-specular intensity following a
well defined curvature. In the R$+-$ map the additional line of
intensity appears at lower exit angles than the specular beam,
while in the R$-+$ map this line is located at higher exit angles
than the specular beam. Moreover, both off-specular intensity
lines exhibit a banana shape curvature, which, in addition, is
intensity modulated in a fashion similar to the specular spin-flip
reflectivity. These two features can be explained qualitatively
and quantitatively, as being due to domain wall scattering.\\
\section*{Discussions }

First we would like to discuss the key elements which define the
intensity of spin-flip scattering at  domain walls (DW).  For one
large domain wall $D_{DW}$, the neutron will adiabatically follow
the magnetic induction and no SF scattering occurs. In contrast,
for very small $D_{DW}$ the spin-flip probability is very small.
Therefore, the thickness of the domain wall affects the scattered
off-specular intensity.  Second, the transmitted intensity through
the DW is affected by the angle $\gamma$ between the magnetization
vectors in neighboring magnetic domains. For $\gamma=0$ no SF
scattering is expected, for $\gamma=90^{\circ}$ the SF scattering
should have a maximum. Also, for $\gamma=180^{\circ}$ one should
observe off-specular scattering because, even so the neutron spin
do not flips, the magnetic field flips instead, leading to non-SF
(in respect to the incident neutron polarization) off-specular
signal. Third, the number of domain walls that the neutron passes
through before being reflected or transmitted by the next
interface affects the off-specular intensity. This gives
information about the average lateral size of the magnetic
domains. Numerical calculations of the transmission coefficient as
function of $D_{DW}$ show that for SF scattering at domain walls a
finite DW width and an angle $\gamma > 0$ is required.\\

The type of the domain wall, Bloch versus N\'eel, can be inferred
from the splitting of the neutron resonances. It was shown that
stray fields emerging from the sample as reported in Ref.
\cite{welp:2003} cause a splitting of the resonant angles for the
different polarizations \cite{radu:2003}. Assuming that Bloch
walls give rise to stray fields, and N\'eel walls do not, then
through the stray fields one can discriminate between these two
types of domain walls. For the present case we have seen no
splitting of the neutron resonances (data not shown). Therefore,
we suggest that at least partial  N\'eel walls are present in the
sample.

Next we discuss the banana shaped curvature of the off-specular SF
scattering. Imagine that down polarized neutrons ($-$) polarized
opposite to the field direction) enter into a magnetic domain as
schematically indicated in Fig.\ref{fig9}. We assume that the
magnetic induction vector in the domain is tilted with respect to
the polarization direction of the incident neutrons as shown in
Fig.\ref{ripple} and Fig.\ref{fig9}. Let us follow the path (red
line) of the neutrons, which are off-specularly scattered into the
$(-+)$ channel. The ($-$) neutrons will meet the vacuum/sample
interface at position \#1 and will then be transmitted without
spin flip into the layer~\footnote{a fraction of neutrons will be
specularly reflected back into the air, and another fraction will
be spin-flipped.}. Their quantization axis will turn parallel to
the magnetic induction inside of the magnetic domain
($\bf{B}_{D1}$) (white shade). Travelling further deep into the
layer, they will be reflected at the bottom interface without
spin-flip. After reflection, they will cross a perpendicular  (to
the surface plane)  magnetic domain wall (\#2) and enter into the
next magnetic domain (grey shade). The orientation of the magnetic
induction inside of this domain ($\bf{B}_{D2}$) is tilted away
from the magnetic induction ($\bf{B}_{D1}$) in the previous domain by a tilt angle $\gamma$. \\
Now we make a basic and fundamental assumption, which is essential
for the interpretation of the banana shaped off-specular
intensity. We assume that the neutrons, which cross a
perpendicular domain wall, are being spin-flipped and reflected
off-specularly. The spin-flip cross-section depends on the tilt
angle $\gamma$. Due to the Zeeman energy change, the SF neutrons
will be refracted and they will follow a different path with
respect to the NSF neutrons~\footnote{Certainly, a fraction of
neutrons will not be spin-flipped. In this case there is no Zeeman
energy change and thus there will be no deviation from their
specular path direction.}. Next, these neutrons will meet the
interface at position \#3 and will exit into the air at an angle
$\theta^{DW-SF}_f$. The same qualitative process also applies for
neutrons transmitted through the substrate
or for $(+)$ neutrons entering the sample from the top (not shown here).\\
The qualitative discussion about the neutron paths in the magnetic
layer can be cast into a set of equations for the reflection and
refraction effects, using Snell's law for neutron optics at each
interface \# 1, 2, and 3 as indicated in Fig.~\ref{fig9}, yielding
the following system of equations:
\begin{figure}[!ht]
    \begin{center}
\includegraphics[clip=true,keepaspectratio=true,width=1\linewidth]{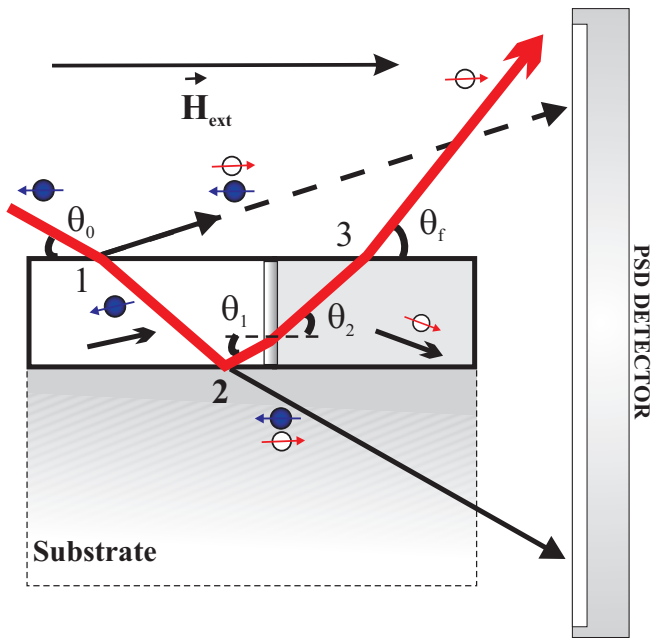}
    \end{center}
        \caption{\label{fig9}Schematic outline of the polarized neutron path in a magnetic
film containing domain walls. The thin red arrows indicate
neutrons oriented parallel to the magnetic field induction inside
and outside of the magnetic film, while thin blue arrows indicate
neutrons oriented anti-parallel to the magnetic field induction.
Thick red line shows the (-+) neutron path before and after
entering the film, which contains two magnetic domains, separated
by a perpendicular magnetic domain wall. The magnetization within
domains is lying in the sample plane. In this graph it is assumed
that  spin flip  occurs only once, at the domain wall where the
neutron enters the grey shade magnetic domain. At the lower
interface the neutron is reflected. At the domain wall between the
white and grey domain, refraction occurs with spin flip. The
neutron leaves the top interface without spin flip. The thin black
lines show neutrons specularly reflected from the top interface
and transmitted neutrons through the lower one. Notice that in
general the neutron beam will meet a variable number of domain
walls, depending on the incidence angle. This will affect the
off-specular intensity but not the position of the streaks.}
\end{figure}
\begin{equation}
\begin{array}{cc}
k_0^\pm\cos(\theta_0)=k_1^\pm \cos(\theta_1^\pm)& \\
k_1^\pm\sin(\theta_1^\pm)=k_2^\pm \sin(\theta_2^\pm)& \\
k_2^\pm\cos(\theta_2^\pm)=k_0^\pm \cos(\theta_f^\pm)&\\
\end{array},
\label{eq1}
\end{equation}
where $k_i^\pm=\sqrt{k_0^2-(u\, \pm |\mu| |B_D|)}$ are the proper
values of the wave vector in the magnetic domains, and
$\theta_i^\pm$ are the angles of the (+) and (-)~neutrons,
respectively, as shown in the Fig.~\ref{fig9}. The solutions of
this system of equations are:
\begin{equation}
\begin{array}{c}
\cos(\theta_f^{DW-NSF})=\cos(\theta_0)\\
\cos(\theta_f^{DW-SF})=\sqrt{{\pm {2\mu B_D}\over(\frac{\hbar^2}{2m}k_0^2)}+\cos^2(\theta_0)},\\
\label{eq2}
\end{array}
\end{equation}
where $\theta_f^{DW-NSF}$ and $\theta_f^{DW-SF}$ are the exit
angles of the neutrons with and without spin-flip upon crossing
the domain wall, respectively, and $k_0= 2 \pi/\lambda$ is the
incident wave number. The formulae above describe very well all
geometric features related to the banana shaped off-specularly
reflected neutrons as observed experimentally. For example, the
$(+-)$ off-specular SF intensity appears only when the incident
angle satisfies the condition $cos^2(\theta_0)\ge {2\mu
B_D}/(\frac{\hbar^2}{2m} k_0^2)$, while for the $(-+)$ curve there
is no limiting incident angle: the equation $cos^2(\theta_0)\ge
{-2\mu B_D}/(\frac{\hbar^2}{2m} k_0^2)$ is always satisfied. This
is the case for $(-)$ neutrons incident on Co, which has a
negative scattering length density. The streaks can be reproduced
by the following set of value (shown by black solid lines in Fig.
\ref{offspecEVA1} ): $|\bf{B}_{D1}|$=$|\bf{B}_{D2}|$= 15500~Gauss,
$\gamma = 30{^\circ}$, and perpendicular domain walls.
$|\bf{B}_{D1,2}|$ is reduced compared to the measured saturation
value of 17~200~Gauss, which we believe it is due to spin canting
within the domains. Furthermore, the intensity of the streaks
contains additional information, from which the average domain
wall width \cite{schaerpf88}, the average domain size
{$D_D~\approx~14~000$ nm}, and the distribution of domain wall
angles can be determined. A full quantitative account of all
features observed in the maps and also a discussion of other
magnetic materials, which have a positive potential barrier
for both incident neutron states, is provided elsewhere~\cite{raduexp}. \\
In passing we stress that the discussion of the off-specular and
asymmetric streaks in terms of optical paths taken by neutrons
inside the sample, is equivalent to geometrical optics of
electromagnetic waves, which explains the propagation of wave
fields.
Clearly, a complete analysis using the distorted wave
approximation should explain both, the geometry and the
intensity~\cite{tope2001}.  There, however, the refraction at the
domain
 walls is not accounted for. The present
analysis has the advantage of providing a clear and lucid
geometric interpretation with quantifiable parameters about the
magnetic induction in the Co film even at remanence.\\

\section*{Conclusions}
In summary, we have measured with polarized neutron reflectivity
the specular and off-specular magnetic scattering from a
ferromagnetic Co film in the remanent magnetic domain state. We
have analyzed the off-specular magnetic neutron scattering using a
$^3$He gas spin-filter.  The off-specular spin-flip scattering is
dominated by streaks above and below the specular ridge. They
appear due to spin-flip transmission at domain walls. These
streaks at well defined angles are clearly to be distinguished
from magnetic diffuse scattering centered symmetrically at the
specular ridge, the latter one being due to the size distribution
of magnetic domains. In the former case we could show that the
streaks can be explained by refraction and reflection of spin
polarized neutrons at perpendicular domain walls. We suggest that
this is a new
method for determining the magnetization M(T,H=0) in thin films with domains.\\
\section*{Acknowledgments}
We would like to thank V. K. Ignatovich for valuable suggestions
and critical reading of the manuscript, and Andrew Wildes and
Boris Toperverg for valuable discussions. This work was supported
by Sonderforschungsbereiche 491 "Magnetische Heteroschichten:
Struktur und elektronischer Transport" of the Deutsche
Forschungsgemeinschaft. The neutron scattering experiments were
performed at the ADAM (BMBF grant No.~03ZA6BC1) and EVA (of Max
Plank Institute, Stuttgart~\cite{EVA}) reflectometers at the
Institut Laue-Langevin, Grenoble, EU.

\end{document}